\documentclass[sigconf, review=false, anonymous=false]{acmart}
\renewcommand\footnotetextcopyrightpermission[1]{}
\usepackage[noend]{algorithmic}
\usepackage[linesnumbered,ruled,noend]{algorithm2e}
\usepackage{multirow,multicol}
\usepackage{comment}

\newcommand{\Paragraph}[1]{~\vspace*{-0.9\baselineskip}\\{\bf #1}}
\newcommand{\nop}[1]{}
\acmConference[SC '25]{The International Conference for High Performance
	Computing, Networking, Storage, and Analysis}{Nov 16-Nov 21, 2025}{St. louis, Missouri}
	\usepackage{url}

\begin{document}
\title{SLO-Aware Scheduling for Large Language Model Inferences}

\author{Jinqi Huang}
\affiliation{%
  \institution{Huawei Technologies Co., Ltd}
}
\email{huangjinqi1@huawei.com}

\author{Yi Xiong}
\affiliation{%
	\institution{Huawei Technologies Co., Ltd}
}
\email{xiongyi26@huawei.com}

\author{Xuebing Yu}
\affiliation{%
	\institution{Huawei Technologies Co., Ltd}
}
\email{yuxuebing3@huawei.com}

\author{Wenjie Huang}
\affiliation{%
	\institution{Huawei Technologies Co., Ltd}
}
\email{huangwenjie16@huawei.com}

\author{Entong Li}
\affiliation{%
	\institution{Huawei Technologies Co., Ltd}
}
\email{lientong@huawei.com}

\author{Li Zeng}
\affiliation{%
	\institution{Huawei Technologies Co., Ltd}
}
\email{zengli43@huawei.com}

\author{Xin Chen}
\affiliation{%
  \institution{Huawei Technologies Co., Ltd}
}
\email{chenxin@huawei.com}
\renewcommand{\shortauthors}{Jinqi Huang et al.}

\begin{abstract}
Large language models (LLMs) have revolutionized applications such as code completion, chatbots, and online classification. To elevate user experiences, service level objectives (SLOs) serve as crucial benchmarks for assessing inference services capabilities. In practice, an inference service processes multiple types of tasks, each with its own distinct SLO. To ensure satisfactory user experiences, each request's distinct SLOs should be considered in scheduling. However, existing designs lack this consideration, leading to insufficient hardware utility and suboptimal performance.

This paper analyzes scenarios to process tasks with varying SLOs, and introduces a simulated annealing-based scheduler to decide request priority sequence based on a request's SLO, input lengths, and possible output lengths. As the first specialized scheduler for multi-SLO scenarios, this work improves SLO attainment by up to 5$\times$ and reduces average latency by 31.6\% on Python-Code-23k-ShareGPT and ShareGPT\_Vicuna\_unfiltered datasets, compared to current state-of-the-art framework vLLM and a new framework LMDeploy.

\end{abstract}

\maketitle

\section{Introduction}
The rapid revolution of LLMs has been witnessed since the debut of ChatGPT \cite{chatgpt}, with a succession of LLMs being unveiled, including Llama \cite{llama} by Meta, Gemini \cite{gemini} by Google, and Qwen \cite{qwen2025qwen25technicalreport} by Alibaba. Since then, LLMs have gained extensive adoption across a multitude of industry applications such as fault maintainence, customer service, recommendation system, code completion \cite{code1, code2}, question answering \cite{qa1, qa2}, knowledge extraction \cite{knowledgeextract1, knowledgeextract2}, document summarization \cite{summarization1, summarization2}, search engines \cite{search1, search2}, etc. 

\begin{figure*}
	\includegraphics[width=1.0\textwidth]{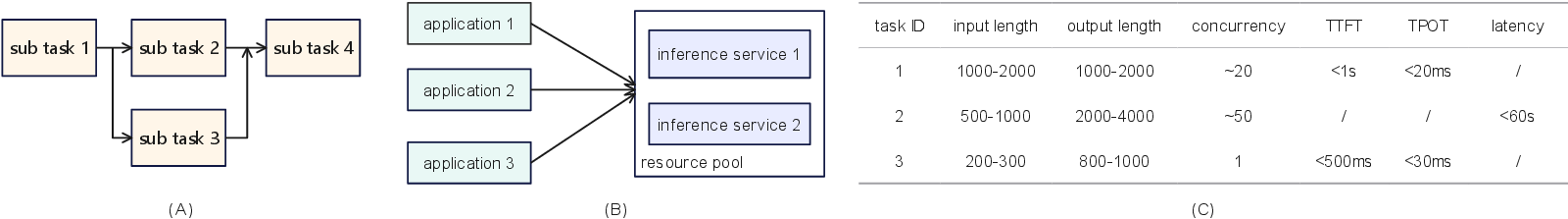}
	\vspace{-0.3cm}
	\caption{Illustration of multiple SLOs existing in LLM inference: (A) distinct SLOs for multiple sub-tasks within an application; (B) diverse SLOs for applications sharing resources; (C) multiple SLOs introduced by scenarios (A) and (B), with `/' indicates unimportance for this task.}
	\label{fig:multi-slo}
\end{figure*}

LLMs are integrated into inference serving systems to provide services. These systems, such as vLLM \cite{vllm}, fastTransformer \cite{fasttransformer}, and Triton \cite{triton}, queue, schedule, and batch the incoming queries before feeding them into LLMs. 
SLOs are introduced to evaluate if a LLM inference service deliver a satisfactory user experience  due to LLM services' interactive nature. 
As the result of diverse user experiences across tasks, SLOs typically vary by applications. 
For an instance, offline processing emphasizes throughput over latency to reduce costs, while online classifiers require real-time responses, making the end-to-end (e2e) latency critical. For chatbots, focusing on time-to-first-token (TTFT) and time-per-output-token (TPOT) ensures quick initial responses and subsequent token generation with human-like reading speed.

In practical industry applications (e.g., fault maintenance, customer service, or recommendation system), a set of LLM inference services are employed for various tasks, each with its own SLO. 
Figures A and B in \ref{fig:multi-slo} show examples of varying SLO requirements for different tasks.
In Scenario 1, as depicted in Figure \ref{fig:multi-slo}(A), an application may involve multiple subtasks, each with varying input and output lengths, request concurrency and specific SLOs. 
Similarly, as shown in Figure \ref{fig:multi-slo}(B), Scenario 2 presents a case where multiple applications with diverse SLOs utilize a shared resource pool, which encompasses more than one LLM inference service. 
Figure \ref{fig:multi-slo}(C) gives an example of multiple SLOs introduced by Scenario 1 or 2. 
To fully utilize the hardware resources and to fulfill application SLOs as much as possible, it is important to account for the varying SLOs across different applications.

Mainstream inference services, including vLLMs \cite{vllm}, Triton \cite{triton}, and fastTransformer \cite{fasttransformer}, focus on high-throughput low-latency scheduling with first-come-first-serve (FCFS) policies. 
Besides, other high-performance schedulers enhance throughput with specialized techniques: Sarathi-Serve \cite{chunkedprefill} introduces chunked prefill to avoid stalls during transitions between prefill and decode phases; FastServe \cite{fastserve} implements skip-join multi-level feedback queues and iteration-level preemption to mitigate head-of-line blocking caused by longer sequences processed before shorter ones; Splitwise \cite{splitwise} and DistServe \cite{distserve} isolate prefill and decode phases into distinct inference instances and optimizing them separately; Llumnix \cite{llumnix} leverages load balancing, prioritization, de-fragmentation, and auto-scaling scheduling strategies to achieve high throughput. 
All these existing designs are tailored to scenarios where queries originate from a single task or from various tasks with identical SLOs. When immigrated to multi-SLO scenarios, these existing systems lack awareness of the unique SLOs associated with different tasks, leading to suboptimal hardware utilization and SLO attainment.

To address the complexity of multi-SLO scheduling, this paper propose a SLO-aware priority mapping algorithm. 
Aiming at optimizing the system SLO attainment and minimizing the average latency, we optimize each request's priority and the actual batch size, based on task-specific SLOs, input lengths, and output lengths prediction. 
To mitigate substantial searching overhead for the optimal priority sequence and request batch sizes, we employ a simulated annealing-based SLO-aware scheduling approach with an overhead of ~1ms, while maintaining a similar effectiveness to an exhaustive search-based method. 
To the best of our knowledge, this work is the first system designed specifically to address SLO-aware scheduling challenges. 
To evaluate the system effectiveness, we benchmark the performance against the state-of-the-art framework vLLM and a newly-introduced framework LMDeploy \cite{lmdeploy} with ShareGPT\_Vicuna\_unfiltered \cite{sharegpt} and Python-Code-23k-ShareGPT \cite{python} datasets, and prove that our system boosts the SLO attainment by up to 5$\times$ and reduces the average latency by up to 31.6\%, respectively. 
The contributions of this paper are summarized below:
\begin{itemize}
	\item We propose a priority mapping algorithm for prioritizing each request based on its SLO, input length, and expected output length if available, to achieve highest SLO attainment with average latency as low as possible.
	\item We develop the first specialized simulated annealing-based SLO-aware scheduler with low overhead to tackle the complexities of SLO-aware scheduling.
	\item We demonstrate the scheduling system on Python-Code-23k-ShareGPT and ShareGPT\_Vicuna\_unfiltered datasets and prove that our method outperforms vLLM and LMDeploy in SLO attainment by up to 5$\times$ and reduces average latency by up to 31.6\%.
\end{itemize}
\section{Background} \label{sec:background}

\begin{figure}
	\includegraphics[width=0.95\linewidth]{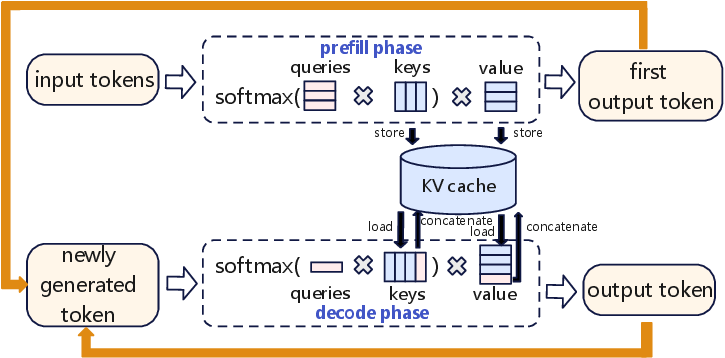}
	\caption{Prefill and decode phases in LLM inferences. KV cache is dynamically stored, loaded, and updated during the inference.}
	\vspace{-0.4cm}
	\label{fig:inference}
\end{figure}

\subsection{Preliminaries}
LLMs evolve from transformers \cite{attention} with an attention-based architecture. Unlike their deep neural networks (DNNs) predecessors, transformer-based LLMs follow the principle of `next token prediction' to construct output sequences: LLM inference generates one token per iteration based on the user prompt and previous tokens.
This working principle forms the characteristic of variable and unpredictable output lengths in LLM. 
Another important feature of LLMs is the significant diversity between prefill and decode phases. 
As shown in Figure \ref{fig:inference}, the prefill phase refers to the procedure from receiving a user prompt to generating the first token. 
During this procedure, tokens from the prompt are processed concurrently to generate a single new token. 
This procedure involves massive matrix multiplications, given that the attention layers of LLMs encompass three core attention matrices—\textit{Queries}, \textit{Keys} and \textit{Values}—all requiring self-attention computations.  
During decoding, the prompt is extended with prior tokens to generate the next one. The \textit{Keys} and \textit{Values} from the previous step are saved as \textit{KV cache} for reuse, with only the \textit{Queries} vector computed from the last token, is employed. 
Therefore, the \textit{KV cache} save computations for decoding from matrix multiplications to matrix-vector products, making data movement the main bottleneck.

\subsection{Related Work}
To achieve high throughput and low latency, numerous inference serving systems provides request scheduling and batching techniques. 
FastTransformer \cite{fasttransformer} utilized request scheduling and static batching technique: requests in the same batch are processed collectively until all have been completed. 
Orca \cite{continuousbatching} was pioneering in introducing continuous batching, a strategy that allows new requests to be seamlessly integrated into the space vacated by completed requests without waiting an entire batch to finish. 
However, due to the inherent differences between the prefill and decode phases, requests in these distinct stages cannot be processed simultaneously within the same batch, causing a stall for prefilling each time when new requests are executed.
Sarathi-serve \cite{chunkedprefill} addressed this challenge by dividing the prefill stage into smaller chunks and interleaving them with ongoing decode operations to form a hybrid batch. 
vLLM \cite{vllm} is the state-of-the-art inference framework that combines both continuous batching and chunked prefill for better flexibility and performance, and employs PagedAttention to eliminates the memory fragmentation. 
LMDeploy \cite{lmdeploy} is a similar newly introduced framework that utilizes model quantization to accelerate LLM inference.
All the aforementioned systems adhere to the FCFS policies, treating all requests equally without consideration for their varying SLO requirements. 

In contrast, FastServe\cite{fastserve} assigns priority to requests based on their prompt lengths and implements skip-join Multi-Level Feedback Queue scheduling along with iteration-level preemption to mitigate the head-of-line blocking inherent in the FCFS and run-to-completion approaches. 
However, FastServe is not specifically tailored for multi-SLO scenarios, as it bases request priorities solely on input lengths. 
Llumnix \cite{llumnix} is another work that takes the priority of requests into consideration. 
It highlighted the needs for new request rescheduling goals such as de-fragmentation and priorities. 
Nonetheless, it fails to provide an executable solution on how to determine the priority of individual requests based on their SLOs. 
SOLA \cite{SOLA} and Tempo \cite{tempo} aim to provide SLO-awareness for scheduling. However, multi-SLO scenarios are not considered.

To fill the vacancy of the multi-SLO scheduler without reinventing the wheel, this paper introduces a decoupled scheduler that can seamlessly integrate with prior historical inference serving systems such as vLLM. 
Other inference serving designs, such as Splitwise \cite{splitwise} and DistServe \cite{distserve}, which focus on maximizing throughput during the prefill and decode phases, as well as solutions like \cite{infinite-llm} and \cite{loongserve} that aim to expedite inference in distributed systems, are orthogonal to the objectives of this paper.

\section{Motivations} \label{sec:motivations}
\begin{figure*}
	\includegraphics[width=1.0\linewidth]{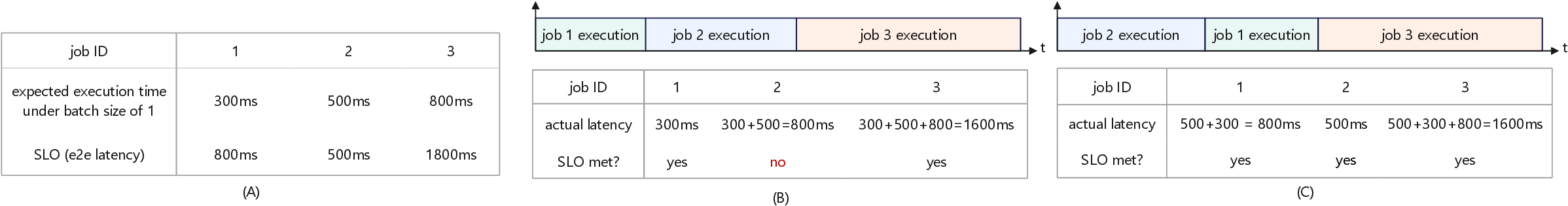}
	\vspace{-0.5cm}
	\caption{Illustration of an SLO-aware scheduling with the batch size of 1 for simplicity:
		(A) Three requests with distinct SLOs are received simultaneously, with their SLO is defined by e2e latency.
		(B) Execution order is based on expected execution time, leading to 2 out of 3 requests meeting SLOs and an accumulated latency of 2700ms, yielding the $G$ of 0.74 req/s.	
		(C) SLO-aware sequencing prioritizes Job 2 over Job 1, with all requests meeting SLOs and an accumulated latency of 2900ms, improving $G$ to 1.03 req/s.}
	\vspace{+0.3cm}
	\label{fig:example}
\end{figure*}
\begin{figure*}
	\includegraphics[width=1.0\linewidth]{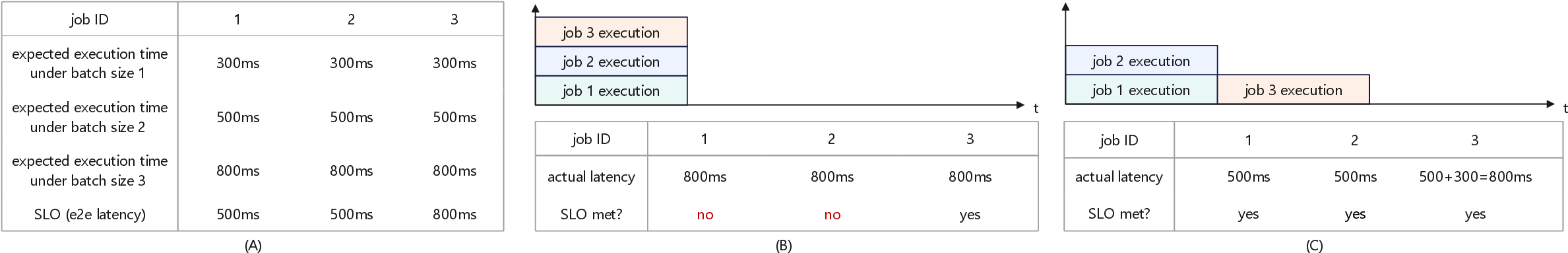}
	\vspace{-0.5cm}
	\caption{Demonstration of how adjusting the batch size enhances $G$. With a max batch size of 3: (A) Three requests with different SLOs are received, and execution time increases with batch size. (B) Processing all three requests in one batch results in only 1 meeting SLOs, with an accumulated of 2400ms, giving $G$ = 0.0004 req/s. (C) Delaying one request to the next cycle allows all 3 to meet SLOs, with an accumulated latency of 1800ms, increasing $G$ to 0.00167 req/s.	}
	\vspace{+0.3cm}
	\label{fig:example3}
\end{figure*}
\begin{figure*}
	\includegraphics[width=1.0\linewidth]{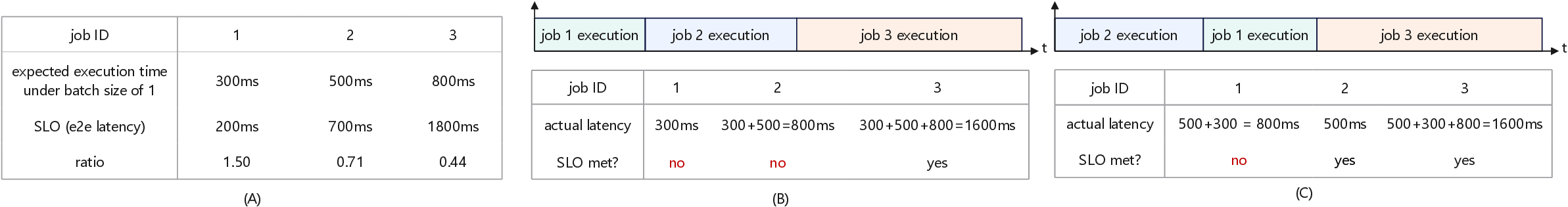}
	\vspace{-0.5cm}
	\caption{Demonstration of deferring a `strict' SLO request boosting $G$ with the batch size of 1 for simplicity: (A) Three requests using e2e latency as the metric are received, including one with an unachievable SLO. (B) Execution prioritizes based on SLO values and the ratio of the expected latency to the actual SLO, with a higher ratio signifying a tougher SLO to meet. Only 1 out of 3 request meets SLO with an accumulated latency of 2700ms, giving $G$ of 0.37 req/s. (C) Delaying the strict SLO allows others to take precedence, with 2 out of 3 meeting SLOs and an accumulated latency of 2900ms, improving $G$ to 0.74 req/s.}
	\label{fig:example2}
\end{figure*}
\vspace{-0.1cm}
\subsection{Problem Formulation} \label{sec:problem}
Traditionally defined goodput is the number of requests that meet SLOs within a time window, as expressed below:
\begin{equation}
	\text{Goodput} = \frac{n}{t_{'}}
\end{equation}

Here, $n$ denotes the number of requests that meet their SLOs, and $t^{'}$ signifies the total latency incurred by processing all requests. This definition frames SLO attainment, the ratio of requests that meet SLOs, as a key metric indicative of user experience. However, $t^{'}$ ignores waiting time’s effect on user satisfaction, with total latency often reflecting the final request’s e2e latency, despite Ref \cite{wait} showing waiting times significantly affect user experience. Moreover, FastServe \cite{fastserve} notes waiting times can be $8\times$ execution time in practice. Furthermore, Ref \cite{goodput} criticizes using goodput as an optimization target, as it can lead to request abandonment, increasing user dissatisfaction over mere SLO misses. To more accurately capture the essence of user experience, we introduce a novel metric $G$ defined as follows. Our objective in this study is to maximize $G$: 
\begin{equation}\label{eq:objective}
	G = \frac{n}{t} 
\end{equation}
Where $t$ denotes the latency summation of all given requests:
\begin{equation}
	t = \sum_{i=0}^{N-1} t_{e2e, i}
\end{equation}

Given a set of job ${j_0, j_1, ..., j_{N-1}}$ with $N$ denotes the number of jobs, their execution time and waiting time are represented as ${t_{exec}^0,  t_{exec}^1, ..., t_{exec}^{N-1}}$ and ${t_{wait}^0,  t_{wait}^1, ..., t_{wait}^{N-1}}$, respectively. The e2e latency experienced by a user can be formulated as follows:
\begin{equation}
	t_{e2e, i} = t_{exec, i} + t_{wait, i}
\end{equation}

Where $i$ is the job index ranging from 0 to $N-1$. In this paper, we consider two streaming service tasks: one prioritizes e2e latency, seen in code completion; another emphasizes interaction speed via TTFT and TPOT, seen in chatbots. $h_i$ indicates which type of tasks $j_i$ belongs to. Offline applications, where throughput is the primary optimization goal rather than user experience, are not in the scope of this work.
\begin{equation}\label{eq:slo}
	h_i=\left\{\begin{matrix}
		1,  & \ \text{if $j_i$ prioritizes e2e latency} \\
		0,  & \ \text{otherwise}
	\end{matrix}\right.
\end{equation}

Given $t_{SLO-e2e, i}$ denotes the SLO (e2e latency), $n$ can be given by the expression as follows:
\begin{equation}
	n = \sum_{i=0}^{N-1} x_i
	\label{eq:slo}
\end{equation}

Here, $x_i$ is defined as a flag to indicate where the job has met the SLO:
\begin{equation}\label{eq:slo}
	x_i=\left\{\begin{matrix}
		1,  & \ \text{if}\ h_i = 1\ \text{and}\ t_{e2e, i} \leq t_{SLO-e2e, i}\\
		1,  & \ \text{if}\ h_i = 0\ \text{and}\ t_{TTFT, i} \leq t_{SLO-TTFT, i}\\
		& \ \text{and}\ t_{TPOT, i} \leq t_{SLO-TPOT, i}\\
		0,  & \ \text{otherwise}
	\end{matrix}\right.
\end{equation}

Where $t_{TTFT, i}$ and $t_{TPOT, i}$can be expressed as below:
\begin{equation}
	t_{TTFT, i} = t_{prefill, i} + t_{wait, i} 
\end{equation}
\begin{equation}
	t_{TPOT, i} = t_{decode, i}
\end{equation}

It is worth noting that altering the job execution order affects both $n$ and $t$ through $t_{wait, i}$. Additionally, it is essential to determine the inference batch size. However, this batch size fluctuates due to the diversity in the lengths of individual requests. Thus, the batch size is expressed as a sequence ${b_0, b_1, b_{M-1}}$ where $M$ represents the total number of batches required for the iterative processing of requests, subject to the constraint that $\sum_{i=0}^{M-1}b_i = N$. Given that the position of a job $j_i$ within the sequence is $p_i, p_i \in [0, N-1]$, the batch index to which $j_i$ is allocated can be denoted as $a_i$:
\begin{equation}
	a_i = K+1,\text{if\ }\sum_{k=0}^{K}b_k \leq p_i < \sum_{k=0}^{K+1}b_k
\end{equation}

Therefore, $t_{wait, i}$ can be expressed as follows:
\begin{equation}
	t_{wait, i}=\left\{\begin{matrix}
		\sum_{j=0}^{a_i-1} \max_{i=0}^{N-1}(t_{exec, i}\cdot g(i, j)) & \ if\ a_{i} \leq 1\\
		0,  & \ otherwise 
	\end{matrix}\right.
\end{equation}

Where $g(i, j)$ is a function to indicate whether job $j_i$ is assigned to batch $j$:
\begin{equation}
	g(i, j)=\left\{\begin{matrix}
		1,  & \ if\ a_i = j\\
		0,  & \ otherwise
	\end{matrix}\right.
\end{equation}

Therefore, the primary objective is to identify a permutation of ${p_0, p_1, ..., p_{N-1}}$ and the batch index of each request ${a_0, a_1, ..., a_{N-1}}$ that maximize $G$ :
\begin{equation}
	\underset{\{p_0, p_1, ..., p_{N-1}, a_0, a_1, ..., a_{N-1}\}}{argmax} G
\end{equation}
Notably, $G$ can also be expressed as the ratio of SLO attainment to average latency. Hence, the primary goal of this paper is to maximize SLO attainment while concurrently minimizing average latency to its lowest feasible level. 
In the following context, we use $G$ to indicate the optimization objective. 

\subsection{Opportunity: Priority Scheduling with SLO Awareness}

The main issue in current inference systems, as outlined in Section \ref{sec:problem}, is the lack of SLO awareness: existing designs either overlook priority scheduling or not consider the varied SLOs on different request types. Figure \ref{fig:example} illustrates an example of SLO-aware scheduling. When three jobs are received concurrently (shown in Figure \ref{fig:example} (A)), historical scheduling methods focused on reducing wait times based on input size or execution time, ignoring individual request SLOs. The strategy in Figure \ref{fig:example} (B) may not efficiently optimize $G$ for requests with different SLOs. To order requests more effectively, priority scheduling with SLO awareness can be applied: Upon receiving requests, their priority is set based on SLOs and predicted latencies using input lengths and predicted output lengths, then executed accordingly in inference systems. Figure \ref{fig:example} (C) demonstrates the SLO-aware scheduling approach that yields a superior $G$. However, integrating ppriority scheduling with SLO awareness into existing inference systems still faces two major challenges.

\vspace{-0.3cm}
\subsection{Challenge 1: Complexity of SLO-Priority Mapping}
SLO-aware scheduling aims to map SLOs and request attributes for processing order. However, constructing such a function is not a simple task. To begin with, various applications require different types of SLOs, including e2e latency, TTFT, TPOT, etc. Meanwhile, different applications exhibit varying degrees of sensitivity to these SLOs. Therefore, it is essential to standardize the expression of primary SLO constraints for each request. 

Additionally, the batch size serves as a tunable parameter affecting determination of the priority sequence through its impact on latency. Figure \ref{fig:example3} shows the effect of batch size changes on $G$. `Strict' SLOs for jobs 1 and 2 (Fig. \ref{fig:example3}A) cause batching all requests to slow down these jobs, despite not exceeding the maximum batch size (Fig. \ref{fig:example3}B). Instead, shifting a `loose' SLO task to the next iteration enhances SLO attainment (Fig. \ref{fig:example3}C), thus improving $G$. Accounting for request batching, determining the SLO-aware priority scheduling sequence is further complicated.

Moreover, prioritizing requests solely by SLOs or the difficulty of meeting SLOs does not always optimize $G$. Occasionally, deferring requests with `strict' SLOs can sometimes improve overall $G$. By allocating the time saved from these requests, one can accommodate a greater number of requests with more easily achievable SLOs, thereby increasing the total count of requests that meet their SLOs. Figure \ref{fig:example2} shows such an example. In Figure \ref{fig:example2} (A), job 1 is unable to meet its SLO. Assigning high priority to job 1 results in job 2 also failing to meet its SLO, as depicted in Figure \ref{fig:example2} (B). However, by slightly delaying the execution of job 1 (shown in Figure \ref{fig:example2} (C)), $G$ is maximized with a cost of slightly increased latency. As a result, these considerations further increases the complexity of the SLO-priority mapping.   
\vspace{-0.3cm}
\subsection{Challenge 2: Inefficient Scheduling Solution Searching}
Searching for the optimal priority order is crucial for better performance, but it is also essential to minimize the overhead, as it contributes to the e2e latency.
Given $n$ requests, an exhaustive search to find the sequence that maximizes $G$ have a time complexity of O($n$!), and traversing over all possible batch size options for each request have a time complexity of O($2^n$), resulting in a total time complexity of O($n! \cdot 2^n$). 
As $n$ grows, the search space explodes, making the search time excessively long (over 2 hours for $n>30$).
Consequently, an efficient search algorithm is key for designing an effective SLO-aware scheduler.

\section{Methodology} \label{sec:methodology}
\subsection{Design Overview}
\begin{figure}
	\includegraphics[width=1.0\linewidth]{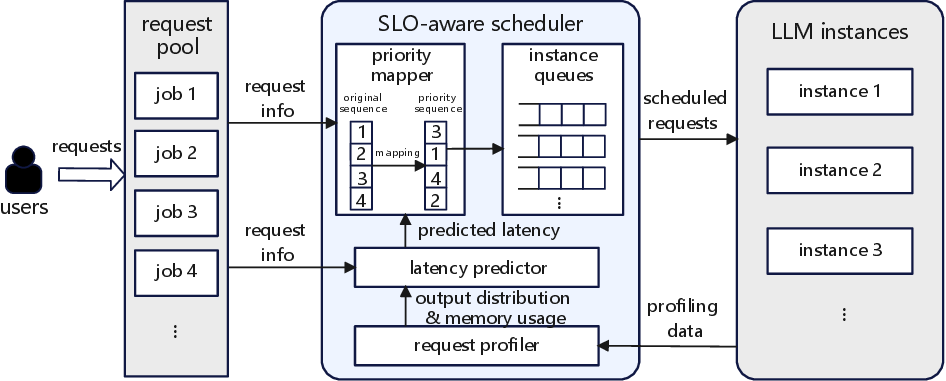}
	\caption{Design overview of the SLO-aware scheduler.}
	\label{fig:overview}
\end{figure}

Figure \ref{fig:overview} presents an architectural overview of the SLO-aware scheduler, which comprises four key components: a request profiler, a latency predictor, a priority mapper, and instance queues. User requests are collected in a request pool, where they await processing. These requests are then fed to the latency predictor and priority mapper for further analysis. The latency predictor combines request information with output distribution and memory usage data provided by the request profiler to predict the latency of each request. Subsequently, the priority mapper leverages the SLOs of the requests along with their predicted latencies to determine the priority order. This priority sequence is then allocated to instance queues, the number of which corresponds to the actual count of LLM inference instances. Finally, requests are dispatched to these instances for execution. 

The implementation details of the request profiler and the latency predictor are elaborated in Section \ref{sec:predictor}, and the core algorithm of the priority mapper and the corresponding scheduling strategy are detailed in Sections \ref{sec:priority} and \ref{sec:solution}, respectively.  

\subsection{Latency Predictor} \label{sec:predictor}
The latency predictor aims to provide a reasonable execution time prediction, which can be further used as a criteria to decide the priority sequence. Following the estimation modeling provided by history work \cite{slsc}, the prefill time $t_{p}(b, l_i)$ increases linearly to both the input length $l_i$ and the batch size $b$ when $l_i$ is below 2k:
\begin{equation}
	t_{p}(b, l_i) = \alpha_p \cdot b \cdot l_i + \beta_p \cdot b + \gamma_p \cdot l_i + \delta_p\label{eq:p}
\end{equation}
Where $\alpha_p, \beta_p, \gamma_p, \delta_p$ denote as fitting coefficiences for prefill time. Likewise, the per-token decode time $\tau_{d}(b, l)$ is linear to both the accumulated length $l_a$ and the batch size $b$ when $l_a$ is below 2k:
\begin{equation}
	\tau_{d}(b, l_a) = \alpha_d \cdot b \cdot l_a + \beta_d \cdot b + \gamma_d \cdot l_a + \delta_d\label{eq:d}
\end{equation}
Where $\alpha_d, \beta_d, \gamma_d, \delta_d$ denote as fitting coefficients for decode time. 
The decode time $t_{d}(b, l_a)$ can be expressed as the accumulation of the per-token decode time, with dependency to the batch size $b$, the input length $l_i$, and the output length $l_o$:
\begin{equation}
	t_{d}(b, l_i, l_o) = \Sigma_{k=1}^{l_o}\tau_{d}(b, l_i + k) 
\end{equation}

Therefore, three typical SLOs (e2e latency, TTFT, and TPOT), without considering the waiting time, can be expressed as follows:
\begin{equation} 
	t_{exec}(b, l_i, l_o) = t_{p}(b, l_i) + t_{d}(b, l_i, l_o)\label{eq:e2e}
\end{equation} 
\begin{equation} 
	t_{prefill}(b, l_i) = t_{p}(b, l_i)\label{eq:prefill}
\end{equation}
\begin{equation} 
	t_{decode}(b, l_i, l_o) = \frac{t_{d}(b, l_i, l_o)}{l_o} \label{eq:decode}
\end{equation}

The SLO-aware scheduler employs a request profiler to gather data across a range of $b$, $l_i$ and $l_o$. This collected data is instrumental in calculating the precise fitting coefficients: $\alpha_p, \beta_p, \gamma_p, \delta_p, \alpha_d, \beta_d, \gamma_d$, and $\delta_d$. Equations \ref{eq:p} and \ref{eq:d} depict multiple linear regression models that include interaction terms. To determine the coefficients, a set of data with diverse values of $l_i$, $l_o$ and $b$ is sampled. The objective is to identify the coefficient set that minimizes the mean squared error between the predicted and actual values of $t_p$ and $t_d$, and coefficient set can be solved using the least squares method.

The remaining inquiries now are: 1) how to ascertain the output length $l_o$, and 2) how to determine the optimal batch size $b$? For the first question, the SLO-aware scheduler gathers data on the task types of incoming requests and offers an optional input variable to allow business users to specify a typical output range or distribution for each task type. If the output range or the distribution is unavailable, the request profiler is equipped with the capability to gather and model the output range and distribution for various task types. This data can subsequently be leveraged to ascertain a probable typical output length, as well as the boundaries for output length. Moreover, established designs, such as S3 \cite{s3} and Ref \cite{responselength}, offer precise predictions of output lengths. These can be integrated with the SLO-aware scheduler to enhance the accuracy of latency predictions. 

For the second question, as depicted in Equations \ref{eq:e2e}, \ref{eq:prefill}, and \ref{eq:decode}, the e2e/prefill/decode latencies are influenced by the batch size, which in turn affects the establishment of the priority sequence. Once the priority sequence is determined, the batch size also needs to be updated accordingly restricted by the remaining memory size and the preset maximum batch size. Consequently, the batch size and the priority sequence cannot be optimized separately. The method of determining the optimal batch size $b$, will be elaborated upon in Sections \ref{sec:priority} and \ref{sec:solution}.

\begin{algorithm}
	\small
	\caption{Simulated-Annealing Priority Mapping}
	\label{alg:mapper}
	\KwIn{Newly arrived jobs $J_{in}$ with valid predicted latency $J_{in}.predE2E$, $J_{in}.predTTFT$, $J_{in}.predTPOT$, simulated annealing parameters initial temperature $T_0$, threshold temperature $T_{thres}$, iteration times $iter$, temperature decay rate $\tau$}
	\KwOut{ Job priority $job.prio, job \in J_{in}$}
	\textbf{procedure} priorityMapping  \\
	\textbf{Initialization:} $T = T_0$ \\ 
	$J_{prio} = sorted(J_{in}, key=J_{in}.predE2E)$ \\
	$f = calculateG(J_{prio})$\\
	$k = 0$\\
	\textcolor{blue}{// check if the sequence with smallest accumulated latency have all requests meet SLOs}\\
	\If{$meetSLONum(J_{prio}) == len(J_{prio})$}
	{
		\For{$job \in J_{prio}$}
		{
			$job.prio$ = job.index($J_{prio}$)
		}
		return	
	}
	\Else
	{
		$f_{init} = calculateG(J_{in})$ \\ 
		\textcolor{blue}{// find the better start point from the sequence with smallest accumulated latency and the initial sequence}\\	
		\If{$f < f_{init}$}
		{
			$J_{prio} = J_{in}$
		}		
	}
	
	\Repeat{$T < T_{thres}$}
	{
		\Repeat{$k == iter$}
		{
			$k += 1$ \\
			\textcolor{blue}{// choose a way to generate a new solution}\\
			$operation = rand({0, 1, 2})$ \\
			\If{$operation == 0$}{
				$J_{cur} = squeezeLastIter(J_{prio})$ \\
			}
			\ElseIf{$operation == 1$}{
				$J_{cur} = delayNextIter(J_{prio})$ \\
			}
			\Else
			{
				$J_{cur} = randSwapping(J_{prio})$ \\	
			}			
			$f_{new} = calculateG(J_{cur})$\\
			\textcolor{blue}{// determine whether to accept the new solution}\\
			\If{$f_{new} > f$}
			{
				$J_{prio} = J_{cur}$ \\
				$f = f_{new}$ \\
			}
			\ElseIf{$exp(-\frac{f_{new} - f}{T}) < rand(0, 1)$}
			{
				$J_{prio} = J_{cur}$ \\
				$f = f_{new}$ \\			
			}
		}
		$T *= \tau$ \\
	}
	\For{$job \in J_{in}$}
	{
		$job.prio$ = job.index($J_{prio}$)
	}	
\end{algorithm}
\vspace{-0.5cm}
\subsection{Priority Mapping Algorithm} \label{sec:priority}
As detailed in Section \ref{sec:predictor}, the batch sizes and the priority sequence must be fine-tuned in a coordinated fashion. Consequently, within the priority mapper, the fundamental strategy for optimizing both parameters involves fixing one (either the batch sizes or the priority sequence) while optimizing the other, and then iterating this process to achieve a refined optimization.

According to the definition of $G$, its upper boundary is achieved if all requests successfully meet their SLOs when the priority sequence is perfectly aligned with the requests’ e2e latency order. In this scenario, the latency summation of all requests $t$ is the smallest with the largest achievable $n$. If not all requests achieve their SLOs when the priority sequence matches the requests’ e2e latency, the priority order that maximizes the objective $G$ needs to be searched.

Given a predefined batch size, input request length, and a potential output length, the latency predictor provides a reasonable estimation of the e2e latency, TTFT, and TPOT, excluding any waiting time. 
Subsequently, with an assumed execution order, the objective function $G$ (as outlined in Equation \ref{eq:objective}) can be calculated. 
Modifying the batch sizes and the execution order yields varying values of $G$. 
By documenting the maximum value of $G$, the optimized batch sizes and the corresponding priority sequence can be ascertained.

\Paragraph{Strawman Solution: exhaustive search}.
To determine the priority sequence and the batch size that yields the highest value for the objective $G$, a straightforward approach involves systematically evaluating every permutation of the request execution order and every possible request batch size combination, calculating the corresponding $G$ for each, and selecting the setting of the batch sizes and the sequence permutation that results in the maximum $G$. Nevertheless, this method of generating all permutations is associated with a time complexity of $O(N!)$, and finding all batch size combination has a time complexity of $O(2^N)$. The total time complexity of traversing over all possible solution is $O(N! \cdot 2^N)$, which can be computationally infeasible when the number of requests $N$ grows. A scheduler employed in practical scenarios cannot accommodate such an overhead, as the scheduling overhead directly contributes to the e2e request processing latency. 

\Paragraph{Our Optimized Solution: Simulated Annealing}. 
An effective strategy for addressing the challenge is to employ simulated annealing to explore sequence permutations and request batch sizes that yield the maximum value of $G$. Simulated annealing, a classic heuristic optimization technique, is renowned for its ability to approximate global optima and has found widespread application across diverse fields \cite{sa1, sa2, sa3}. Drawing inspiration from the process of cooling in materials, this algorithm introduces a random perturbations to the current solution, generating a new solution in the process. The new solution is unconditionally accepted if it improves upon the current score. Conversely, even if the new solution is less effective, it may still be accepted with a probability, thus facilitating the escape from local optima. 

To incorporate simulated annealing into the priority mapping process, the pseudo code is listed in Algorithm \ref{alg:mapper}. 
There are two starting solutions: one is the original input sequence order with all request batch sizes set to the maximum; another is generated by aligning the priority sequence according to the e2e latency of the requests. If the second solution satisfies the SLOs for all requests, the search is terminated, and the optimal solution is deemed to have been discovered. If not, the one starting solution that has higher $G$ is set as the start point, and the simulated annealing algorithm commences its search: the temperature $T$ decays, and at each temperature level the process iterates $iter$ times to approximate the optimal solution. During this phase, the algorithm has three possible ways to create a new sequence: squeezing a request into the last batch iteration if the current request is not in the first iteration and the last iteration batch size has not reach the maximum; delaying a request into the next iteration if the next iteration batch size has not reach the maximum; randomly exchanging the positions of two requests in the priority sequence. If the objective function $G$ for the new sequence exceeds that of the previous sequence, it is adopted. Alternatively, even if the new sequence yields a lower $G$ value, it may still be accepted based on a probability that is dependent on the current temperature $T$: the higher the temperature, the greater the likelihood of accepting a less optimal solution. The search terminates once the temperature falls below the predefined threshold. 

The overall time complexity of the search process is $O(t \cdot iter)$, with $t$ represents the number of steps required for the temperature to decrease to the threshold level. Significantly, the computational time for the simulated annealing search does not scale linearly with the number of requests. Nonetheless, to preserve the effectiveness, it is imperative to adjust the hyperparameters $T$ and $iter$ upwards when the number of requests increases, as this expansion accommodates a broader search space.

\begin{algorithm}
	\small
	\caption{SLO-Aware Scheduling}
	\label{alg:scheduler}
	\KwIn{Newly arrived jobs $J_{in}$, remaining memory per instance $M_0, M_1, ..., M_{InstNum-1}$, and instances queues ${Q0, Q1, Q_{InstNum}}$}
	\KwOut{Jobs to be executed per instance $J_{out, 0}, J_{out, 1}, ..., J_{out,InstNum-1}$ for one iteration}
	\textbf{procedure} SLOAwareScheduling \\
	\textbf{Initialization:} $J_{out, 0}, J_{out, 1}, ..., J_{out,InstNum-1} = 0, 0, ..., 0 $\\
	\textcolor{blue}{// instance pre-assignment; job latencies are predicted in this stage}\\
	$J_{in, 0}, J_{in, 1}, ..., J_{in,InstNum-1} = InstAssign(J_{in})$ \\
	\For{$J_{cur} \in \{J_{in, 0}, J_{in, 1}, ..., J_{in,InstNum-1}\}$}
	{	
		\textcolor{blue}{// use the latency prediction and the SLOs to decide priority} \\
		$J_{cur} = priorityMapping(J_{cur})$ \\
		$J_{cur} = sorted(J_{cur}, key=J_{cur}.priority)$ \\
		\textcolor{blue}{// instance assignment}\\
		\For{$job \in J_{cur}$}
		{
			$Q_{cur}.push(job)$\\
		}
	}	
	
	\textcolor{blue}{// schedule jobs to execute} \\
	\For{$Q \in \{Q_0, Q_1, ..., Q_{InstNum}\}$}
	{
		$inst = Q.index$ \\
		$J_{out, inst} = ScheduleReq(J_{in}, Q, M_{inst})$
	}
	\Return $J_{out, 0}, J_{out, 1}, ..., J_{out,InstNum-1}$ \\
\end{algorithm}
\vspace{-0.2cm}
\subsection{Scheduling Solution} \label{sec:solution}


The application of SLO-aware scheduling can be expanded to encompass multiple instances as needed. Consequently, the scheduling solution encompasses both instance assignment and request scheduling. The instance assignment plays a crucial role in determining which requests are prioritized during the priority mapping phase. Ideally, the instance assignment and priority mapping should be coordinated to maximize the value of $G$ across all requests. However, this approach encounters the challenge of a search space that expands linearly with the number of requests, which can impede efficiency even when employing simulated annealing for the search process, particularly in scenarios with a high volume of concurrent requests. To address this, the scheduling solution first allocates requests to instances and then performs priority mapping within each instance independently. If LLM instances are hosted on different servers, the priority mapping can be carried out in a distributed fashion, thereby further accelerating the process.

The operational flow of the scheduling solution is as follows: initially, requests latencies are predicted, and distributed to instances in a round-robin fashion. Next, the priority sequence is established through the priority mapping algorithm, as described in Section \ref{sec:priority}. Requests are then enqueued to instance-specific queues and scheduled for execution on LLM instances. The pseudo-code for this scheduling process is provided in Algorithm \ref{alg:scheduler}.  \newline
\Paragraph{Instance Assignment}. 
Following the aforementioned workflow, the logic for assigning request sequences to instances is to first predict the latency of the requests, and then allocate requests in a round-robin way, with each request being assigned to the instance possessing the largest remaining memory, thereby achieving load balancing. When the instance with the largest remaining memory lacks sufficient space to accommodate a new instance, the remaining memory is reset, indicating that a maximum possible number of requests have been allocated and a fresh iteration starts. The equation to estimate the number of tokens with a given memory usage is as follows:
\begin{equation}
	token\_num(m) = \frac{m \cdot \mu}{\sigma}
\end{equation}
Where $m$ is the remaining memory, $\mu$ denotes the memory utility that is less than 1 due to memory fragmentation, and $\sigma$ is the memory consumption per token. 
In actual practice, $\mu$ and $\sigma$ need to be ascertained. 
$\mu$ is adjusted by computing the ratio of the peak memory consumption to the available memory prior to executing any inference and then averaging this ratio. 
Meanwhile, $\sigma$ can be derived by dividing the aggregate memory consumption by the total number of tokens processed.

\Paragraph{Request Scheduling}. 
The request scheduling is straightforward: once the final priority sequence is determined and each request is assigned to an instance, the requests are then dispatched into the queue associated with their respective instances. A batch of requests is then scheduled for execution from this queue as soon as the corresponding LLM instance is prepared and available.

\vspace{-0.07cm}
\section{Evaluation} \label{sec:evaluation}

In this section, we evaluated the SLO-aware scheduler through a series of experiments. 
Initially, we showcase the overall performance enhancements over varied framworks (vLLM and LMDeploy), models (Qwen2.5-7B and Qwen2.5-32B), and devices (NVIDIA V100 GPUs and NVIDIA A800 GPUs).    
Next, we delve into the influence exerted by individual features. 
Finally, we test the scalability across instances.
\vspace{-0.2cm}
\begin{figure*}
	\includegraphics[width=1.0\linewidth]{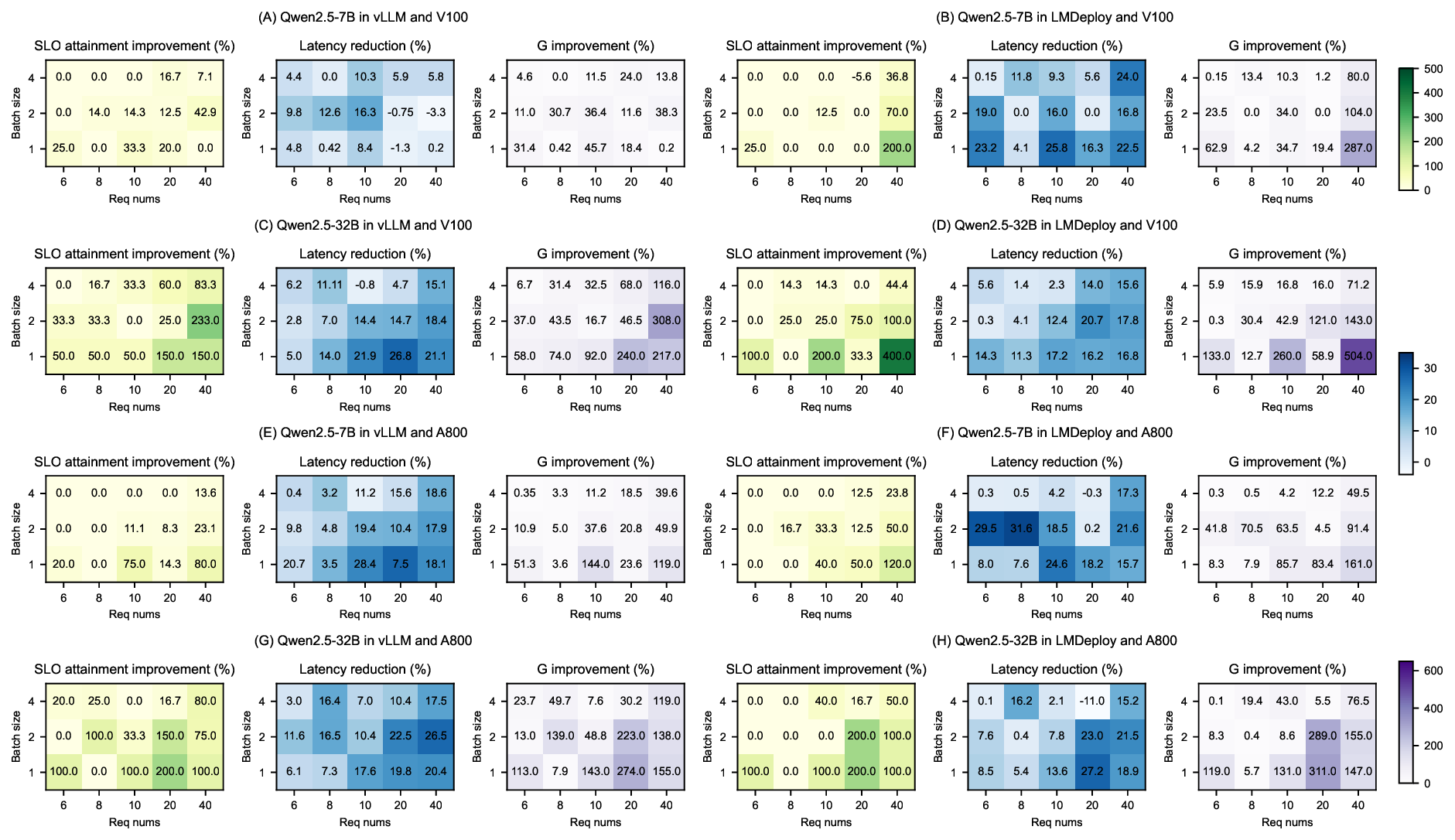}
	\vspace{-0.6cm}
	\caption{Overall performance with (A) Qwen2.5-7B in vLLM in V100, (B) Qwen2.5-7B in LMdeploy in V100, (C) Qwen2.5-32B in vLLM in V100, (D) Qwen2.5-32B in LMdeploy in V100, (E) Qwen2.5-7B in vLLM in A800, (F) Qwen2.5-7B in LMdeploy in A800, (G) Qwen2.5-32B in vLLM in A800, (H) Qwen2.5-32B in LMdeploy in A800, evaluated in a mixed dataset of ShareGPT\_Vicuna\_unfiltered dataset and Python-Code-23k-ShareGPT dataset with varying request numbers and maximum batch sizes. }
	\label{fig:res1}
\end{figure*}
\vspace{-0.2cm}
\subsection{Experimental Settings}
\Paragraph{Testbed}. 
The evaluation utilizes 2 NVIDIA V100 32GB GPUs or 1 NVIDIA A800 80GB GPUs.
The NVIDIA V100 GPUs are connected over PCIe 3.0 x16, equipped with 516GB host memory and 2 Intel Xeon 6132 CPUs. 
The NVIDIA A800 GPU is equipped with 1546GB and 2 Intel Xeon 6240 CPUs. The scalability tests were deployed to up to 8 V100 GPUs in the same server.

\Paragraph{LLM models}. 
The LLM models utilized in the experiment are Qwen2.5-7B and Qwen2.5-32B\cite{qwen2.5}, which have been widely used in industry and academia.
For Qwen2.5-7B, one instance utilizes 2 NVIDIA V100 GPUs, and \textit{FP16} precision is used during the inference. Experiments with Qwen2.5-7B in 1 A800, Qwen2.5-32B in 4 V100s, and Qwen2.5-32B in 1 A800 were also delivered.

\Paragraph{Workloads}. 
The dataset employed for this experimental study is a composite of the ShareGPT\_Vicuna\_unfiltered dataset \cite{vicuna} and the Python-Code-23k-ShareGPT dataset \cite{python}, both of which are parts of the ShareGPT collection \cite{sharegpt}. 
However, they cater to distinct task domains: the former is tailored for chatbot interactions, while the latter is designed for code generation. 
In current experiment, the combined dataset is constructed by evenly selecting requests from each source, with each request being tagged according to its respective task type. 
For the requests sourced from the dataset ShareGPT\_Vicuna\_unfiltered, the evaluation metrics are TTFT and TPOT, whereas for the requests from Python-Code-23k-ShareGPT, the focus is on measuring the e2e latency because a code is useful only when completed. Request lengths in both datasets are restricted to under 2k for the latency predictor's validation.

\Paragraph{SLOs}. 
The SLOs reflect user experiment expectations for a certain task. 
Historical research \cite{slo} suggests establishing the e2e latency SLO at 10$\times$ the duration of processing an individual request. 
Based on profiling data with Qwen2.5-7B in 2 V100 GPUs, given a request of the Python-Code-23k dataset with an average length, the time required to process it is 3s. 
Consequently, the e2e latency SLO is set at 30s. 
For interactive tasks like a chatbot, the TTFT is defined as 10s, and the TPOT is 50 milliseconds. 
In practical applications, SLOs influence model deployment and the maximum allowable batch sizes for LLM services. 
Nevertheless, to demonstrate the impact of the SLO-aware scheduler, we maintain the SLOs for all experiments.

\Paragraph{Evaluation metrics}. 
The metrics for evaluating the experiments encompass the objective function $G$, as delineated in Equation \ref{eq:objective}. 
Furthermore, to explore the impact of SLO-aware scheduling, we have also presented the SLO attainment rates and the average latency metrics. 
In addition, we have quantified the overhead associated with implementing the SLO-aware scheduling mechanism.

\Paragraph{Baselines}. 
The simulated-annealing SLO-aware scheduler operates as a dedicated scheduling component, necessitating its integration within a LLM service. 
For the experimental evaluation, we have adopted vLLM v0.6.0, a state-of-the-art LLM serving system renowned for its support of PageAttention and continuous batching, as our baseline platform. 
The simulated-annealing SLO-aware scheduler is implemented atop this vLLM system. Besides, we also used LMDeploy v0.7.2 to validate the effectiveness in varied framework. 

Moreover, to assess the performance and efficiency of the simulated annealing-based SLO-aware scheduler, we have also conducted an comparative analysis with an exhaustive search-based SLO-aware scheduler integrated within the vLLM framework. 
Due to the inefficient searching of exhaustive searching, the comparison was confined to small batch sizes and a rather limited number of requests.

\Paragraph{Implementations}. 
The experiment script is implemented as 1.7k lines of \textit{Python} code. 
The default values for the hyperparameters are set as follows: $T_0$ to 500, $T_{thres}$ to 20, $iter$ to 100, and $\tau$ to 0.95 unless otherwise specified. 
As suggested by official instructions, the memory utilizing efficiency is set to be 0.9 for vLLM.

\Paragraph{Workflows}. 
The procedure of the experiments was conducted as follows: Initially, following the deployment of a LLM instance, a series of profiling rounds were carried out to gather data on the e2e, prefill, and decode latencies across a range of batch sizes and request lengths. 
For this study, profiling was performed with the maximum batch size varying from 1 to 32, and the request lengths ranging from 100 tokens to 8,000 tokens. 
The fitting coefficients are stored to be used for later latency prediction. 

Next, the mixed dataset was created by randomly sampling an equal number of requests from two datasets: Python-Code-23k-ShareGPT and ShareGPT\_Vicuna\_unfiltered dataset. 
Following this, the dataset was shuffled to randomize the order of the entries. 
To ensure a fair comparison across the baseline, for the exhaustive-search SLO-aware scheduling and the simulated-annealing SLO-aware scheduling, the same random seed was utilized. 
However, the seed was modified when applying different experimental settings to prevent the reuse of the same subset of the dataset.

The output length predictor tracked the actual lengths of the outputs once a request’s response was produced, and dynamically modeled this data using a Gaussian distribution. 
To estimate the output length for a new request, the predictor generates a random integer based on this fitted distribution. 

When the SLO-aware feature was disabled, requests were dispatched to the baseline service continuously, allowing the service to determine the actual batch size for each iteration. 
Conversely, when SLO-aware scheduling was activated, requests were submitted to the baseline service in a predetermined order and batch size: requests within the same batch were sent concurrently, whereas requests in separate batches were separated by a brief interval of 0.1ms to prevent them from being combined into a single batch.
\begin{figure}[]
	\includegraphics[width=1.0\linewidth]{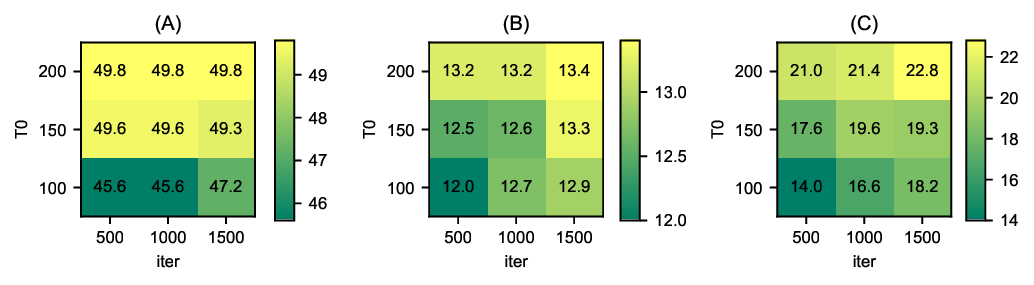}
	\vspace{-0.7cm}
	\caption{Improvement of $G$ with varying $T_0$ and $iter$ in simulated annealing-based priority mapping. (A) involves 10 requests with a maximum batch size of 1, (B) includes 20 requests with a maximum batch size of 2, and (C) features 40 requests with a maximum batch size of 4. }
	\vspace{+0.3cm}
	\label{fig:res2}
\end{figure}
\vspace{-0.05cm}
\subsection{Overall Performance}

\begin{table*}[]
	\begin{tabular}{c|cccc|cccc}
		\hline
		\multirow{2}{*}{req num} & \multicolumn{4}{c|}{SA}                                                                                                        & \multicolumn{4}{c}{Exhaustive search}                                                                                     \\ \cline{2-9} 
		& \multicolumn{1}{c|}{SLO attainment} & \multicolumn{1}{c|}{Latency (s)} & \multicolumn{1}{c|}{G ($\times 10^{-5}$)} & Overhead (s) & \multicolumn{1}{c|}{SLO attainment} & \multicolumn{1}{c|}{Latency (s)} & \multicolumn{1}{c|}{G ($\times 10^{-5}$)}    & Overhead (s) \\ \hline
		4                        & \multicolumn{1}{c|}{1}              & \multicolumn{1}{c|}{10.29}               & \multicolumn{1}{c|}{9.72}      & 0.00023      & \multicolumn{1}{c|}{1}              & \multicolumn{1}{c|}{10.18}                & \multicolumn{1}{c|}{9.82} & 0.0012       \\
		6                        & \multicolumn{1}{c|}{0.83}           & \multicolumn{1}{c|}{11.44}               & \multicolumn{1}{c|}{7.26}      & 0.00032      & \multicolumn{1}{c|}{0.83}           & \multicolumn{1}{c|}{12.00}               & \multicolumn{1}{c|}{6.92} & 0.039        \\
		8                        & \multicolumn{1}{c|}{0.875}          & \multicolumn{1}{c|}{14.48}               & \multicolumn{1}{c|}{6.04}      & 0.00043      & \multicolumn{1}{c|}{0.875}          & \multicolumn{1}{c|}{14.50}               & \multicolumn{1}{c|}{6.03} & 88.2         \\
		10                       & \multicolumn{1}{c|}{0.4}            & \multicolumn{1}{c|}{27.61}               & \multicolumn{1}{c|}{1.45}      & 0.00048      & \multicolumn{1}{c|}{0.4}            & \multicolumn{1}{c|}{28.38}               & \multicolumn{1}{c|}{1.41} & 287          \\ \hline
	\end{tabular}
	\vspace{+0.4cm}
	\caption{Comparison of simulated annealing-based priority mapping and exhaustive search. }
	\vspace{-0.6cm}
\end{table*}
\begin{figure}
	\includegraphics[width=1.02\linewidth]{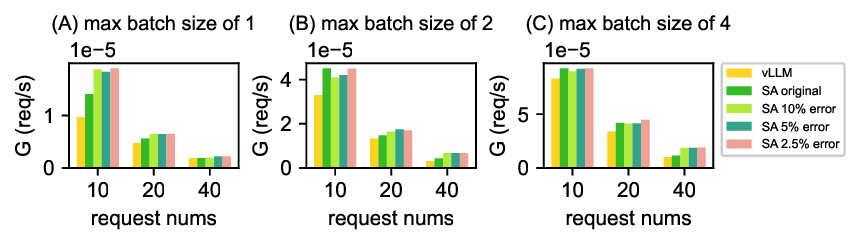}
	\vspace{-0.6cm}
	\caption{Impact of output length prediction on the enhancement in simulated annealing SLO-aware scheduler under the maximum batch sizes of (A) 1, (B) 2, and (C) 4.}
	\vspace{-0.2cm}
	\label{fig:res3}
\end{figure}
We compared the SLO attainment, the average latency, and their ratio (denoted as $G$) across varying request numbers and batch sizes. 
This evaluation was performed using the simulated-annealing SLO-aware scheduler (SA), an exhaustive-search counterpart, and vLLM which lacks SLO-awareness. 
Figure \ref{fig:res1} presents the experimental outcomes for (A) Qwen2.5-7B in vLLM in V100, (B) Qwen2.5-7B in LMdeploy in V100, (C) Qwen2.5-32B in vLLM in V100, (D) Qwen2.5-32B in LMdeploy in V100, (E) Qwen2.5-7B in vLLM in A800, (F) Qwen2.5-7B in LMdeploy in A800, (G) Qwen2.5-32B in vLLM in A800, (H) Qwen2.5-32B in LMdeploy in A800. 

Owing to the variability in request sequences, the enhancement of the metric $G$ spans a range from 0 to 504\%. 
Occasionally, the initial request execution order was sufficiently efficient with minimal waiting times, leading to only modest improvements from the simulated annealing process (as observed in the case with 40 requests and a batch size of 1 in Figure \ref{fig:res1} (A)); or even no improvement (as seen with 8 requests and a batch size of 4 in Figure \ref{fig:res1} (A)), as the result of variations in execution time. 
However, compared with vLLM and LMDeploy, in most cases the simulated-annealing SLO-aware scheduler shows prominent improvement. Among results with varied experiment settings showed growths up to 5$\times$ in SLO attainment (with 40 requests and the maximum batch size of 1 in Figure \ref{fig:res1}(D)) and up to 31.6\% (with 8 requests and the maximum batch size of 2 in Figure \ref{fig:res1}(F)) in average latency. A case with strict SLOs and a large request number (as shown in Figure \ref{fig:res1}(G)-(H)) produced a worse baseline and a larger search space, hence showed a more significant improvement in SLO attainment. On the contrary, the average latency improvement depended on more the baseline sequence randomness than the model or framework performance. Besides, it can also be seen that sometimes the SLO attainment (seen in Figure \ref{fig:res1} (B)) or the latency (seen in Figure \ref{fig:res1} (A), (C), (F) and (H)) is sacrificed to achieve an overall higher $G$.

Furthermore, the simulate annealing-based priority mapping solution demonstrated performance with a maximum degradation of just 1.0\% across the tested examples, compared to the exhaustive counterpart.
We further compared the execution overhead of two priority algorithm with the maximum batch size of 1 and the request number across 4 to 10 and showed the results in Table \ref{tab:overhead}. Results for request numbers exceeding 10 were not recorded, due to prohibitively long computing time from exhaustive search. 
As request numbers increase, the exhaustive search solution's execution time exhibits exponential growth, whereas that of the simulated annealing method demonstrates only a slight increase, which is attributed to the overhead of additional request management. 
This validates the efficiency of our simulated-annealing solution.

We conducted further experiments to investigate the influence of the initial temperature $T_0$ and the number of iterations $iter$ on the improvement of $G$ within our simulated-annealing solution. 
The outcomes are depicted in Figure \ref{fig:res2}. 
Overall, the results indicate that under a specific temperature, increasing the initial temperature $T_0$ yields more significant improvement in $G$ compared to increasing the iteration time $iter$. 
For an instance, in the case with 10 request and a maximum batch size of 1 (as illustrated in Figure \ref{fig:res2} (A)), the improvement of $G$ has grown from 45.6\% to 49.8\% when the initial $T_0$ was elevated from 100 to 200, whereas increasing $iter$ shows an less substantial improvement in $G$ (i.e., from 45.6\% to 47.2\%). Examples with different batch sizes and request numbers depicted in Figure \ref{fig:res2} (B) and (C) also shows this tendency. 
This is because rising the initial temperature $T_0$ not only increments the number of iterations but also heightens the probability of accepting suboptimal solutions, thereby providing more opportunities to escape local optima. 
However, it is not always beneficial to have excessively high values for $T_0$ or $iter$. 
While a higher $T_0$ increases the probability of accepting suboptimal solutions, it also necessitates a correspondingly higher $iter$ to explore more effectively for superior solutions. 
Moreover, a substantial $T_0$ or $iter$ can result in decreased efficiency, as the computational time is directly proportional to these parameters. 
In practice, it is advisable to fine-tune $T_0$ and $iter$ in alignment with the size of the actual search space to optimize performance.
\vspace{-0.2cm}
\subsection{Impact of Output Length Prediction}

We have also explored the effects of output length prediction on our system. Currently, the SLO-aware scheduler utilizes profiling data from tasks to dynamically update the mean and the standard deviation of tasks' actual outputs, which in turn are employed to predict the output lengths of incoming requests. 
The experimental setup was conducted as follows: rather than relying on profiling data to predict output lengths, we employed actual output lengths with variations of 2.5\%, 5\%, and 10\% to simulate the use of an output length predictor with different levels of accuracy. 
Figure \ref{fig:res3} illustrates the impact of the accuracy of output length prediction on the performance enhancement of our SLO-aware scheduler, across different maximum batch sizes: 1 (as shown in Figure \ref{fig:res3} (A)), 2 (as shown in Figure \ref{fig:res3} (B)), and 4 (as depicted in Figure \ref{fig:res3} (C)). 

A discernible trend emerges, indicating that the greater the accuracy of the output length prediction, the more enhancement compared to the current solution, though the inherent randomness of the simulated annealing process still introduces slight variations in the actual outcomes. 
For example, in the scenario with 40 requests and a maximum batch size of 4 (as depicted in Figure \ref{fig:res3} (C)), employing a single output length predictor with an error margin of less than 2.5\% resulted in a 65\% increase in $G$ over the original SLO-aware scheduler, and an 84\% enhancement in $G$ when compared to the baseline. 
While developing a highly accurate output length predictor is beyond the scope of this work, the findings nonetheless showcased the benefits of incorporating an accurate output length predictor with the SLO-aware scheduler in real-world applications. 
These results also motivated us to enhance the predictor’s effectiveness in future research endeavors.
\begin{figure}
	\includegraphics[width=1.0\linewidth]{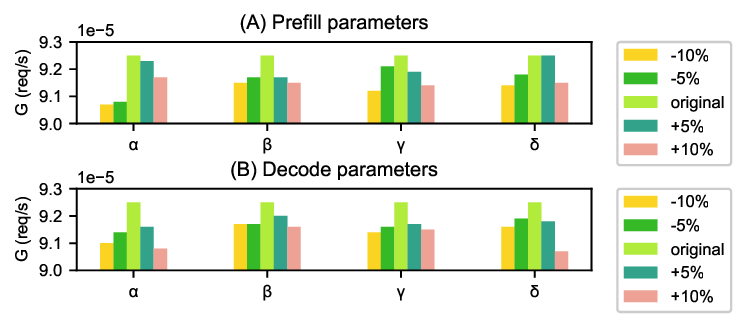}
	\vspace{-0.7cm}
	\caption{Impact of variations on fitting parameters of the latency predictor with 10 requests and the maximum batch size of 4.}
	\vspace{-0.03cm}
	\label{fig:res5}
\end{figure}
\begin{table}[]
	\begin{tabular}{l|lllll}
		\cline{1-5}
		parameter   & $\alpha$&$\beta$  &$\gamma$ & $\delta$ &  \\ \cline{1-5}
		for prefill & 0.1     & 5.7     & 0.01    & 43.67   &  \\ \cline{1-5}
		for decode  & 0.0002  & 0.275  & 0.00088    & 15.85   &  \\ \cline{1-5}
	\end{tabular}
	\vspace{+0.5cm}
	\caption{Fitting parameters used in the latency predictor. }
	\vspace{-1.0cm}
	\label{tab:predict}
\end{table}
\vspace{-0.2cm}
\subsection{Impact of Latency Prediction}

Adhering to the methodology illustrated in Section \ref{sec:predictor}, the fitting parameters derived from the profiling data are presented in Table \ref{tab:predict}. 
Nonetheless, in practical scenarios, factors such as inadequate sampling may lead to variations in these parameters, potentially resulting in the usage of suboptimal fitting parameters. 
To investigate the effects of such variations, we conducted an experiment using a sample scenario with 10 requests and a maximum batch size of 4. 
The results of this experiment are depicted in Figure \ref{fig:res5}. 
Overall, there is a correlation between the degree of deviation from the optimal parameters and the severity of degradation in $G$, with the most significant decline reaching up to 1.9\%. 
Furthermore, variations in $\alpha$ have been identified as the most impactful, as it directly affects the accuracy of estimating the influence of batch size and processing lengths. 
Therefore, it is recommended to collect a sufficient amount of data for parameter fitting to mitigate the effects of variations in the fitting parameters.
\begin{figure}
	\includegraphics[width=1.0\linewidth]{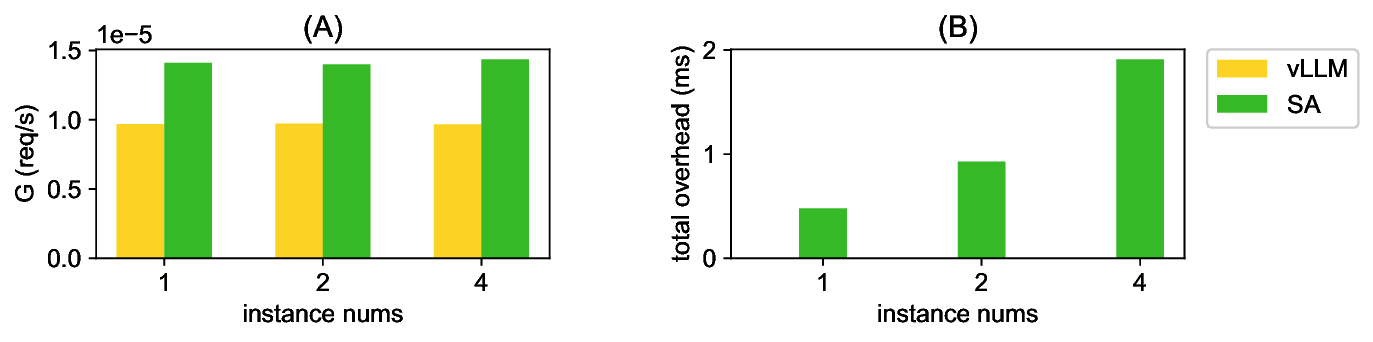}
	\vspace{-0.6cm}
	\caption{The enhancement of $G$ (A) and the overall overhead (B) when applied across multiple instances on the same machine.}
	\vspace{-0.2cm}
	\label{fig:res4}
\end{figure}
\vspace{-0.2cm}
\subsection{Scalability Tests}

We have expanded our experiments to encompass multiple instances. 
Figure \ref{fig:res4} illustrates the impact of the SLO-aware scheduler as the number of instances ranges from 1 to 4. 
In this setup, each instance was allocated to 2 V100 GPUs, with all instances collocated on a single server. 
To mitigate the effects of randomness, 10 requests were dispatched to a single instance, and these requests were then replicated across all instances, with the number of copies matching the number of instances. 
The figure clearly demonstrates that the effectiveness of the SLO-aware scheduler is sustained when scheduling is performed within each instance. 
There is a linear increase in the time taken for scheduling, which rises to 0.93 ms for 2 instances and 1.91 ms for 4 instances. 
This escalation is due to the sequential execution of multiple instances on the same server, which can be accelerated by parallelism in the future.

\vspace{-0.2cm}
\section{Conclusion} \label{sec:conclusion}
In this paper, we introduced a novel priority mapping method and invented the first SLO-aware scheduler designed for scenarios where LLM inference requests exhibit varying SLOs. 
With the goal of optimizing system SLO compliance while maintaining the lowest possible average latency, we computed the priority of each request and determine the optimal batch size for execution based on task-specific SLOs, input lengths, and output lengths prediction. 
Through the simulated annealing-based SLO-aware scheduling, we achieved comparable effectiveness to the exhaustive search-based counterpart while incurring an overhead of merely ~1ms. 
We further evaluated the effectiveness of our system against the cutting-edge schedulers vLLM and LMDeploy using the mixed dataset of Python-Code-23k-ShareGPT and ShareGPT\_Vicuna\_unfiltered dataset, demonstrating that our solution can boost the SLO attainment by up to 5$\times$ and reduce the average latency by up to 31.6\%, respectively. 

\section{Acknowledgment}
We acknowledge ChatGLM for AI-assisted proofreading during the writing of the whole paper.
\bibliographystyle{ACM-Reference-Format}
\bibliography{sample}

\appendix
\section{Experiment Detailed Results}
\begin{figure*}
	\includegraphics[width=1.0\linewidth]{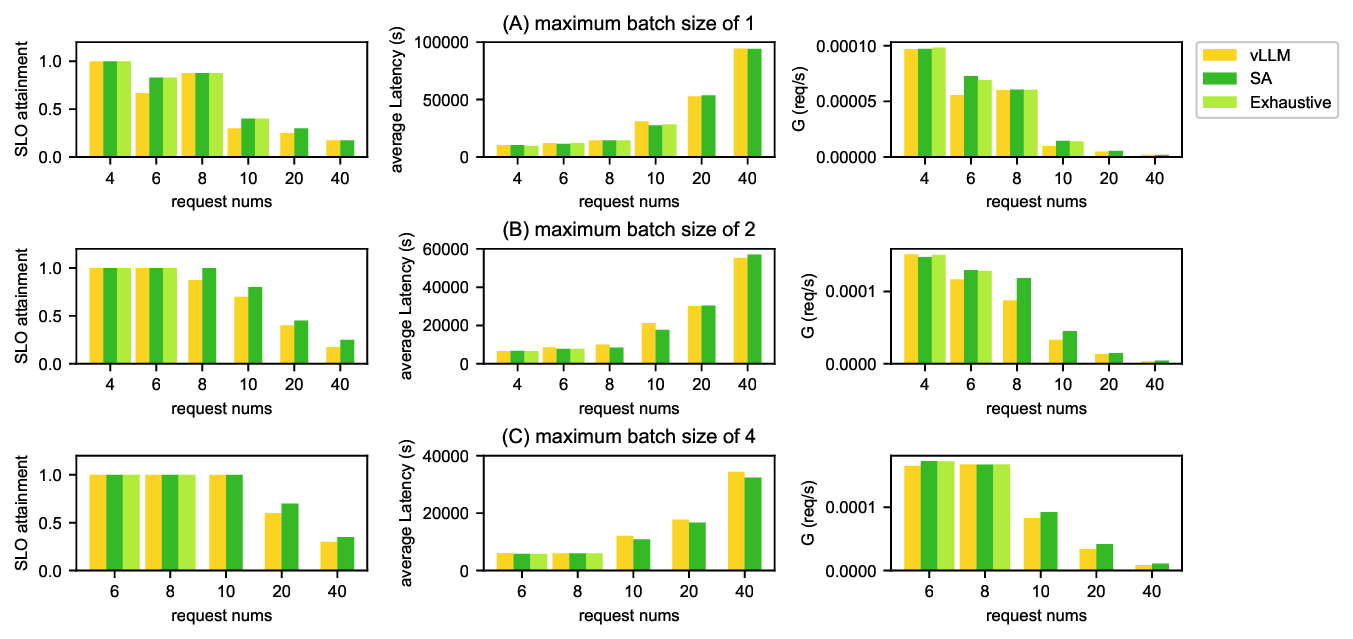}
	\caption{Qwen2.5-7B at 2 Nvidia V100 GPUs in vLLM}
\end{figure*}

\begin{figure*}
	\includegraphics[width=1.0\linewidth]{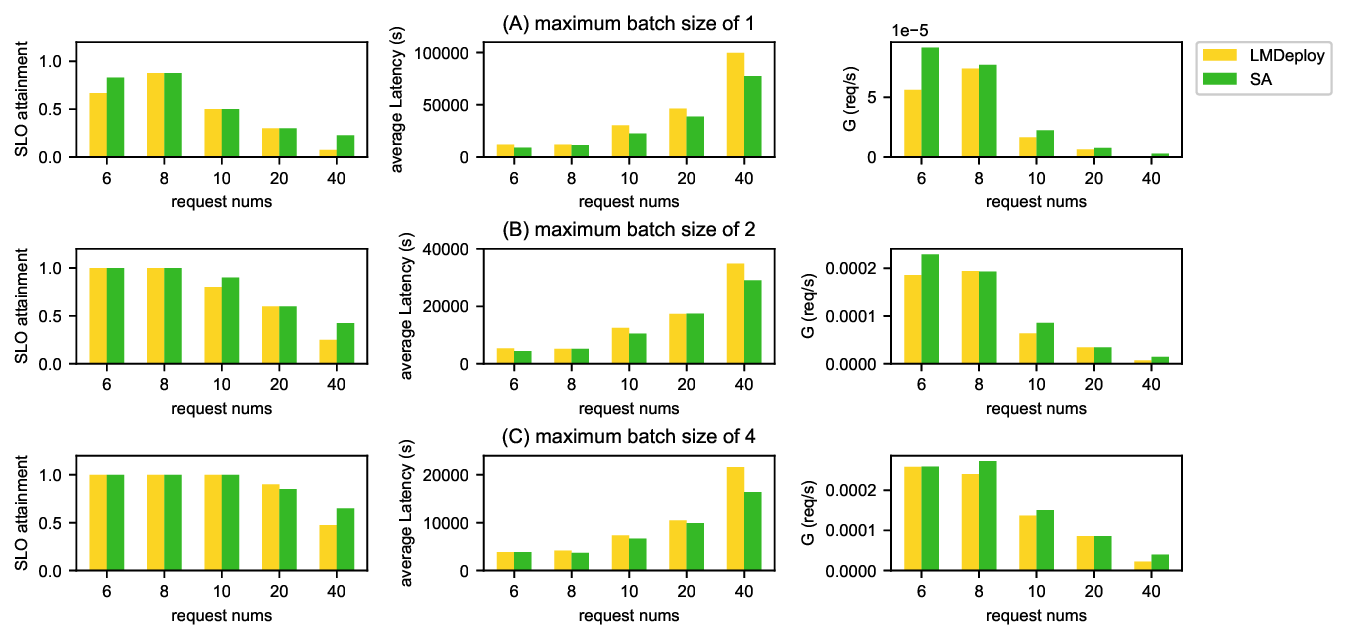}
	\caption{Qwen2.5-7B at 2 Nvidia V100 GPUs in LMDeploy}
\end{figure*}

\begin{figure*}
	\includegraphics[width=1.0\linewidth]{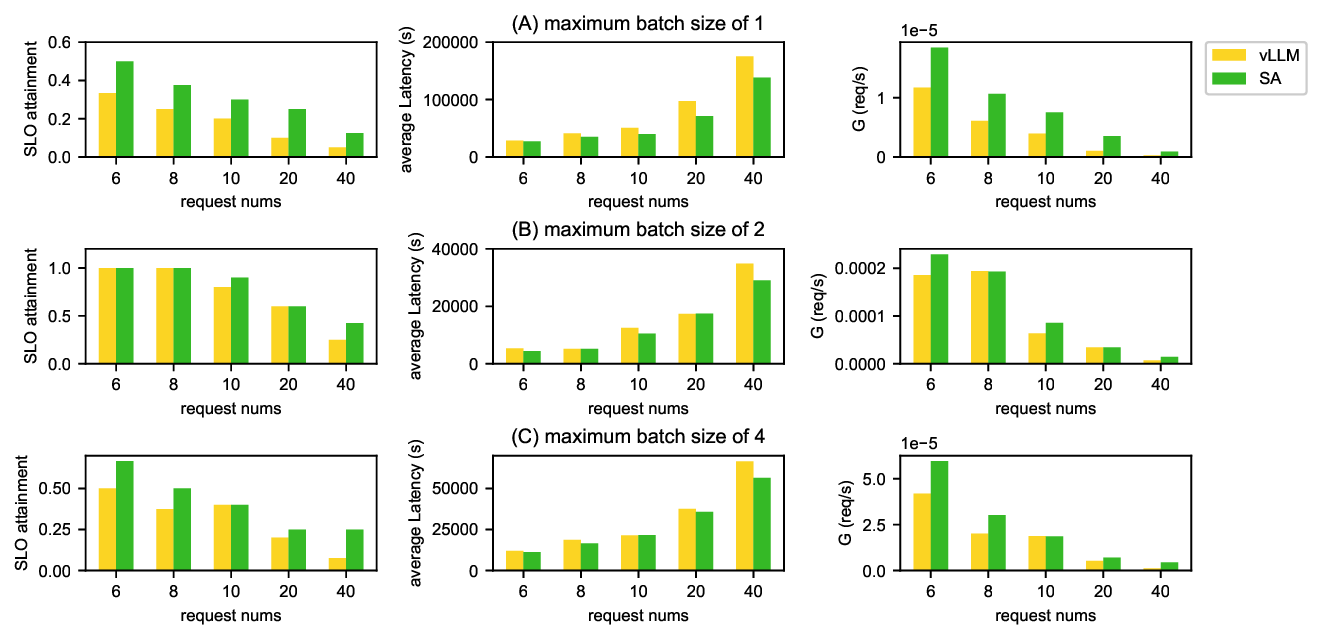}
	\caption{Qwen2.5-32B-Instruct at 4 Nvidia V100 GPUs in vLLM}
\end{figure*}

\begin{figure*}
	\includegraphics[width=1.0\linewidth]{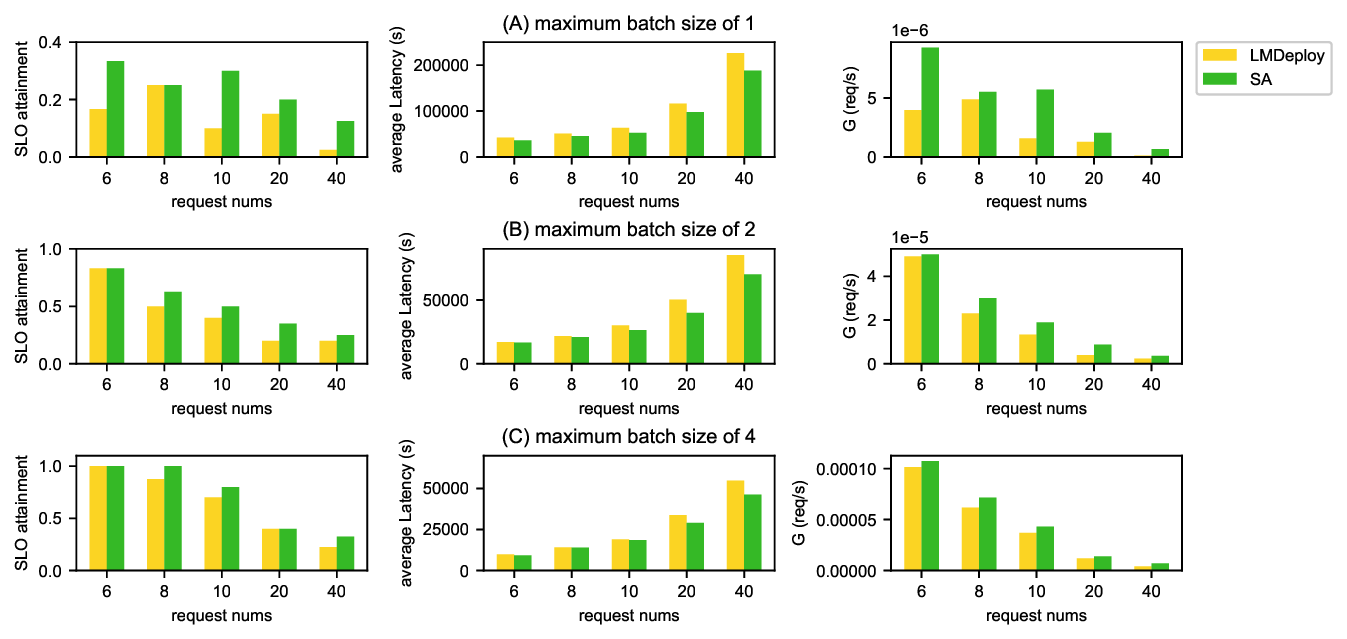}
	\caption{Qwen2.5-32B-Instruct at 4 Nvidia V100 GPUs in LMDepoy}
\end{figure*}

\begin{figure*}
	\includegraphics[width=1.0\linewidth]{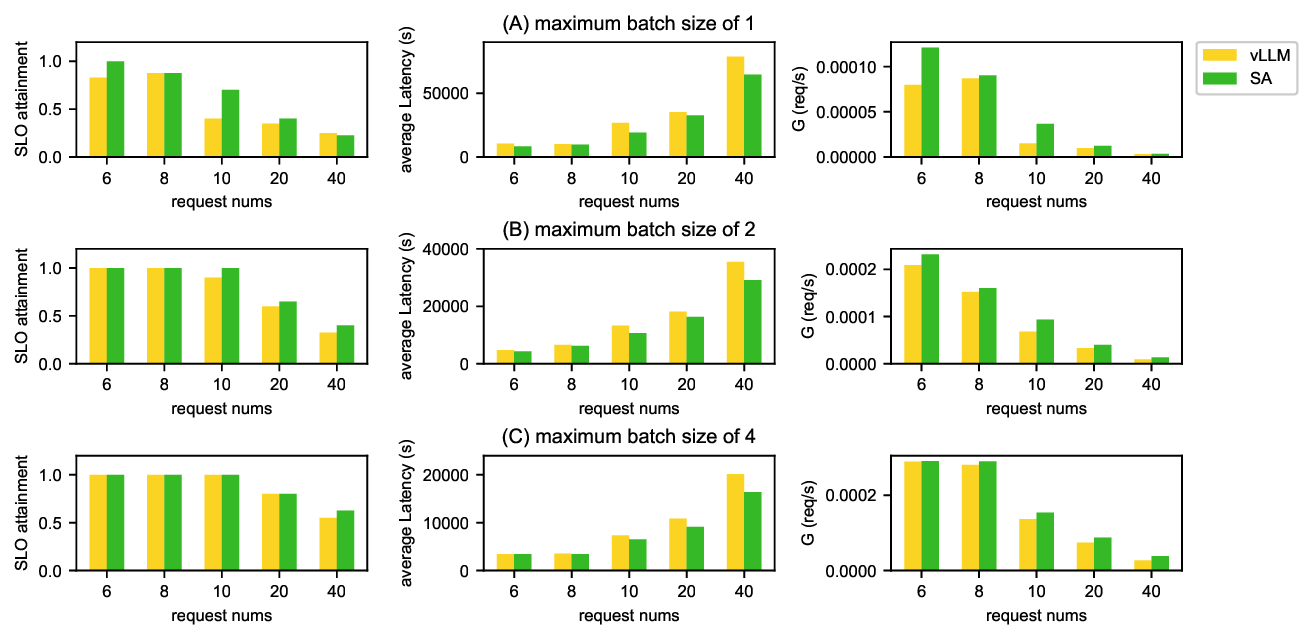}
	\caption{Qwen2.5-7B at 1 Nvidia A800 GPUs in vLLM}
\end{figure*}

\begin{figure*}
	\includegraphics[width=1.0\linewidth]{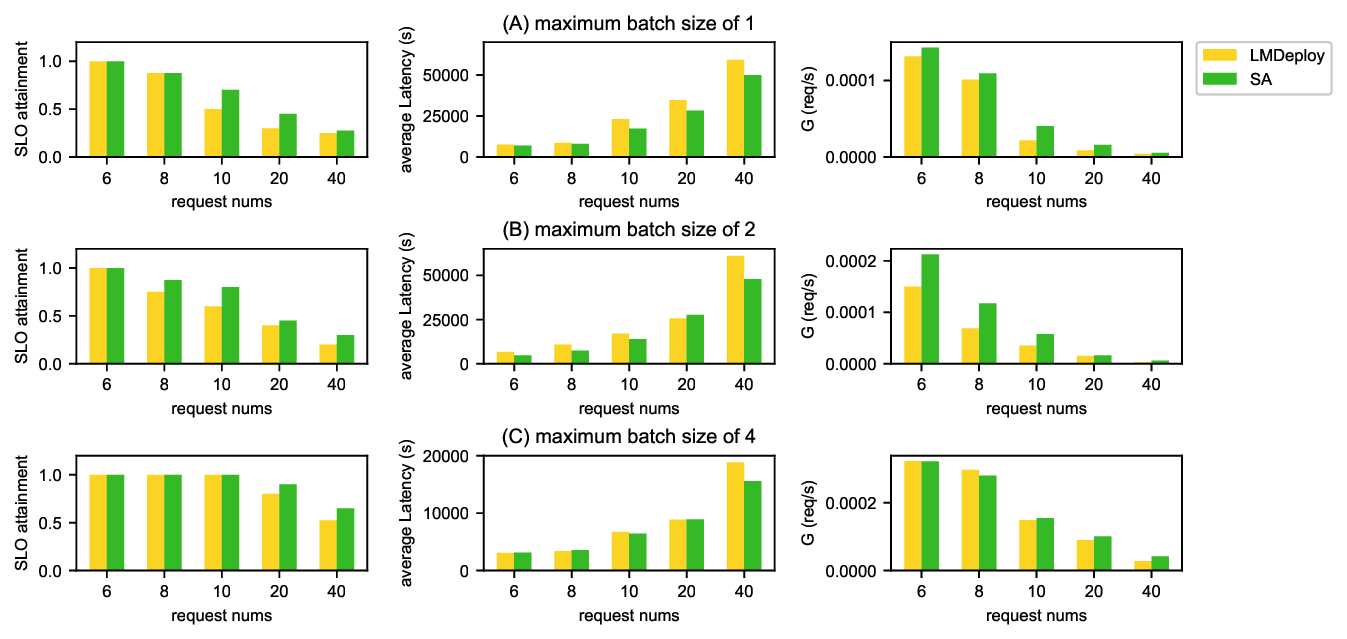}
	\caption{Qwen2.5-7B at 1 Nvidia A800 GPUs in LMDeploy}
\end{figure*}

\begin{figure*}
	\includegraphics[width=1.0\linewidth]{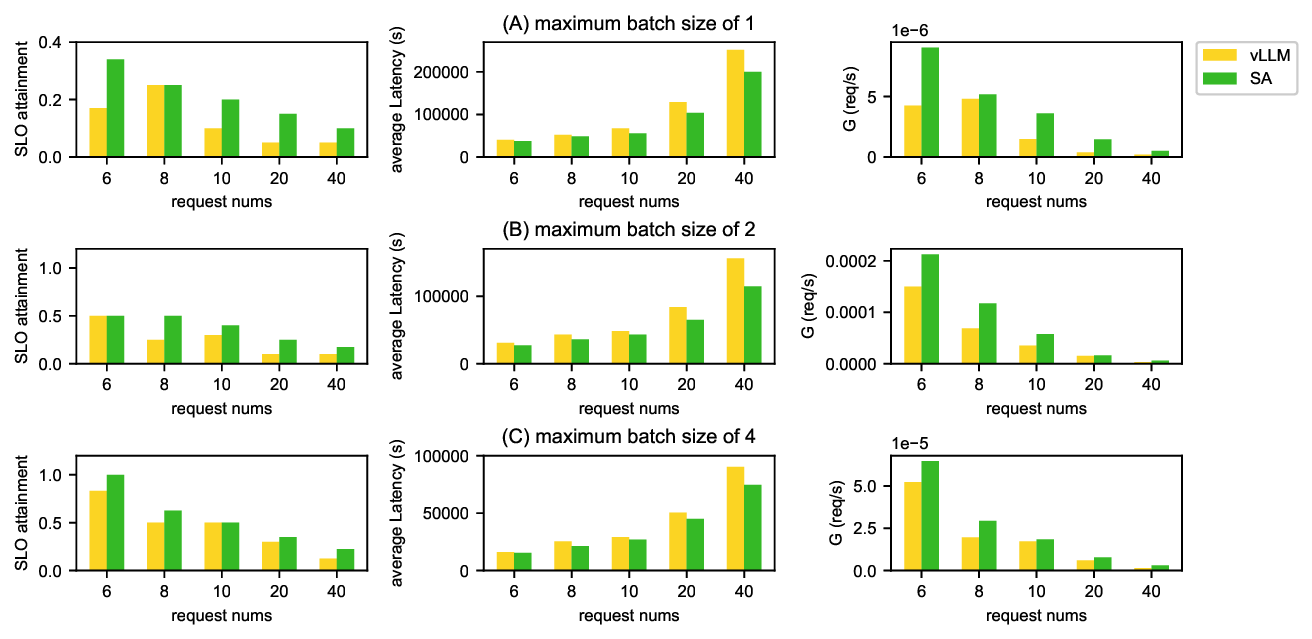}
	\caption{Qwen2.5-32B-Instruct at 1 Nvidia A800 GPUs in vLLM}
\end{figure*}

\begin{figure*}
	\includegraphics[width=1.0\linewidth]{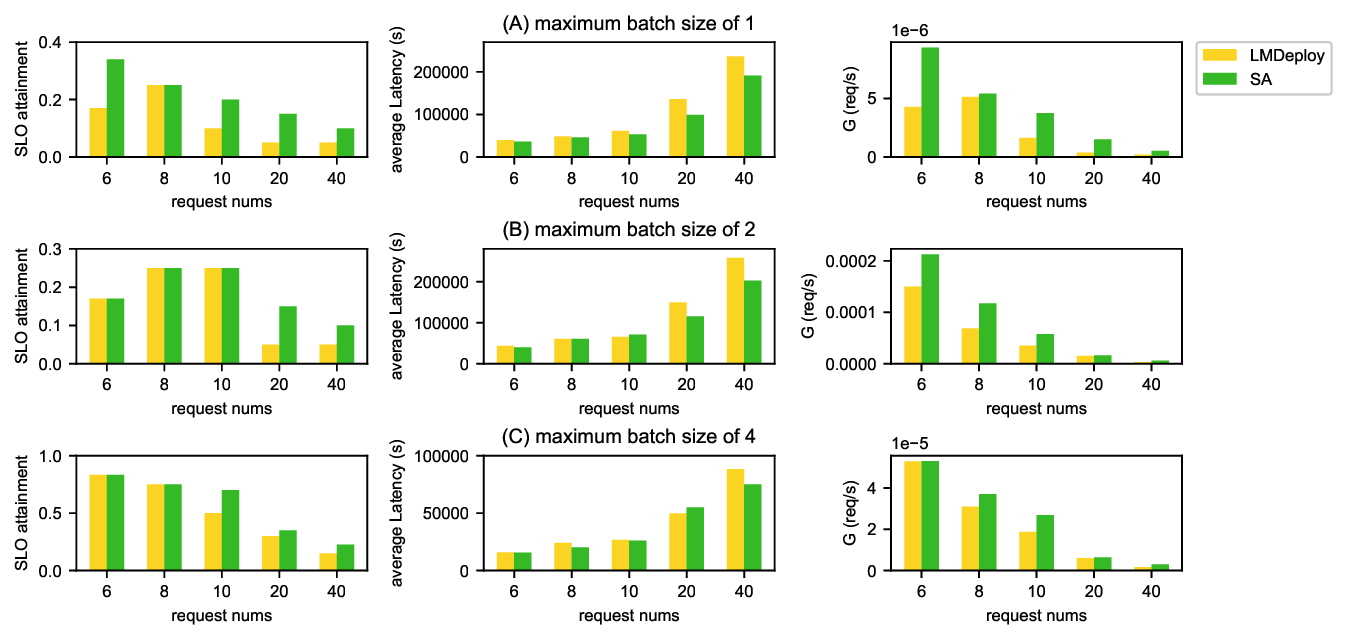}
	\caption{Qwen2.5-32B-Instruct at 1 Nvidia A800 GPUs in LMDeploy}
\end{figure*}

\end{document}